# DESIGN A MULTICULTURAL BLENDED E-LEARNING SYSTEM

Hayder Hbail

Razi University, Department of Computer Engineering

h.jasim@pgs.razi.ac.ir

*Abstract*

Most universities in developing countries are using a teaching and learning approach known as blended e-learning, however, there was no multicultural-based blended e-learning framework. Furthermore, there is no research to show the impact of multicultural blended e-learning on satisfaction of learners. This research employed two categories of students, the Iranian students and the Iraqi students studying at Razi University of Iran. These two groups were taught using the multicultural blended e-learning approach. We utilized an open source application named Claroline. Questionnaires were designed and administered to students to collect information about the level of satisfaction and the optimal mix of tools that go into blended multicultural-based e-learning. The collected information was analyzed using SPSS. We found that blended multicultural based e-learning improves the level of satisfaction of learners. In addition, the optimal mix of multicultural blended learning should be comprised 19% still pictures, audio files 23%, video files 31% and text files 27%.

A. Introduction

With the expansion of information technology and the penetration of the telecommunication equipment, influence on depth of society have also been changed tools and methods of trainings. Tools and methods are developed to facilitate learning at anytime and anywhere for individuals, with their own facilities and any time period they determined. . In recent years distance learning have been discussed, this approach has some pros and cons. Distance learning was initially possible with paper letter. While the advancement of technology and most importantly, low cost use of technology, made it favorable to use newer tools for transferring knowledge. . With the advent of the Internet standards, tools and methods were introduced for e-learning and this field have been developed every day. . In fact, e-learning is the use of conveying information tools such as electronics (e.g., the Internet) to transmit information and knowledge.

The purpose of this research was to describe and analyze the critical factors, which may affect the implementation of multicultural E-learning system. Investigating the barriers of multicultural E-learning system's implementation would help to decrease the barriers in future implementations. Moreover, we aimed to examine the cultural influence of an organization towards E-learning system's implementation process and identify a complete outline for organizations to deal with the barriers for E-learning system implementation process and hence to make the E-learning system work efficiently and successfully.

This study may be inspiring for organizations interested in initiating e-learning systems. . They can find out the tips and methods for implementation of E-learning in their organizations. This study also could be interesting for those organizations that have already implemented E-learning system. Furthermore, this research provides a general guidance for those who are interested in Information System particularly E-learning system. E-learning is a fledgling industry in educational technology and distance learning, but statistics educational institutions, especially universities try to acceralarate the offer of a model to fit the educational structure and culture of the country in the field of e-learning. In addition to the advantages of e-learning one of the main reasons for the need to organize electronic educational institutions in Iran is the growing demand for education, especially higher education in the country. Moreover, due to limited resources and training capacity the growing demand for education is converted to a social issue in the current educational system.

The Effectiveness of e-learning can be removed a problem partially thus according to defined objectives for education and higher education in Iran it is clear that the importance of central and electronic educational institutions in particular in virtual universities in Iran were considered. There is no doubt that the traditional education system in these days could not answer the needs of today's information society, then it is necessary the system was required to be devoted within its configuration varied and we observe adaptability process in the line with the needs of today's societies.

The objectives of the study are:

- To determine whether the use of the blended e-learning based on an LMS approach can bring maximum satisfaction to academic needs of students at universities in developing countries notably the Razi University.

- To establish how best e-learning elements can be blended to come up with the best mix that suits with real differences among the cultures may impede the design and the learning process and learning needs of learners at universities in developing countries notably the Razi University.

- To develop a blended multicultural e-learning framework suitable for the
Razi University

### B. Research background

The need for this type of research is further justified by a quotation from Koper (2003) who averred that, "there is lack of integration and harmonization in the e-learning field and even very basic theories and models about e-learning are missing". The Razi University is one of such institutions where there is lack of blending in the e-learning field. A review of literature shows that there is no suitable e-learning framework for institutions in developing countries such as the Razi University which is the major pivot of this research. Existing frameworks are general and not weighted hence are not suitable for institutions in developing countries.

When it comes to the definition of blended learning, it is worth noting that there is no a universally accepted definition of the phrase, however, Heinze Aleksej (2008) suggested that a better term for blended learning is blended e-learning. The term blended e-learning is often used interchangeably with the term blended learning which refers to the mix of traditional instructor-led training, synchronous online conferencing or training and asynchronous self-paced study (Singh, 2003). Blended multimedia-based e-learning frameworks being used in institutions in developed countries cannot be transferred to institutions in developing countries because of differences in educational settings in developing and developed countries (Chimombo 2005, 131). Hence, there is a knowledge gap which this research seeks to address. It must be noted that there are already many e-learning programs offered in developing countries which were developed by various national and international initiatives. Some of these programs are Aptech and World Links for Development Program (Hawkins, 2002). Most 0f such programs were implemented through international institutions or operate cross-nationally but they note that during the transfer of such programs, country specific differences and differences between learners, notably, learner backgrounds and language should be taken into account since these can be barriers to achieve the objectives of e-learning programs.

From the foregoing, it is evident that either the adaptation of the learning resources or general customization of the entire e-learning system should be considered if e-learning systems are to be transferred from developed to developing countries. To support the above point of reusing existing technologies and contents in different contexts, Richter and Pawlowski (2007, 4528) noted that various types of customizations have to be considered when transferring e-learning systems from developed to developing countries at institutional level. Further to that, Brusilovski (2001, 96) established that, specific needs of users have to be identified to supply adequate adaptation of learning resources.

Chimombo (2005, 131) indicated that educational settings in developing countries are different from settings in developed countries and that these countries are characterized by low quality of education and narrow possibilities in attending schools in rural areas because of far distances and high opportunity costs. Concerning the foregoing issue, Kohn *et al.* (2008) argued that, barriers resulting of no or little customization of eLearning systems can be of cultural character, technological character, and are due to differences in prior knowledge.

Rabahallah and Ouamer in 2015 in an article entitled: "Creating e-Learning Web services rowards reusability of functionalities in creating e-Learning systems "pay attention to the importance of the web in the manufacture of electronic systems and noted that the use of function is effective in e-learning and can be a force for development and automating the creation of training appropriate for the individual student. Those seeking to design an LMS to establish effective training programs have both a challenge and opportunity in overcoming the cultural barriers to a successful implementation.

### C. E-learning

Itmazi and Tmeizeh (2008) claimed that e-learning became a hot topic in the 1990's after the spread of the Internet. Furthermore, they stated that, although e-learning has a relative short history, it is becoming an important part of learning. The majority of the universities adopted some kinds of e-learning within their learning systems. They defined e-learning as any learning that could be realized in a computer connected generally with an Internet or Intranet network.

1. The benefits of e-learning

> The aims of e-learning system are to promote knowledge and skills of human resources with the use of the program to date and affordable. Even if the organization employees have enough time to attend traditional classrooms, still alive learning based on classroom will be costly for the organization. In addition, employees must update their information simultaneously with the progress of technology. Another benefit of e-learning is as follows: Learning routine for everyone and everywhere, saving costs, Cooperation and interaction, learning without fear, the ability to choose different levels.D. Multicultural education:

That is a set of strategies and materials in education that were developed to assist teachers when responding to the many issues created by the rapidly changing demographics of their students. It provides students with knowledge about the histories, cultures, and contributions of diverse groups; it assumes that the future society is **pluralistic**. It draws on insights from a number of different fields, including **ethnic studies** and **women_studies**, but also reinterprets content from related academic disciplines.(Banks,1995) Multicultural education, also viewed as a way of teaching, promotes principles such as inclusion, diversity, democracy, skill acquisition, inquiry, critical thought, value of perspectives, and self-reflection. It encourages students to bring aspects of their cultures into the classroom and thus, allows teachers to support the child's intellectual and social/emotional growth. Multicultural education is also attributed to the reform movement behind the transformation of schools. Transformation in this context requires all variables of the school to be changed, including policies, teachers' attitudes, instructional materials, assessment methods, counseling, and teaching styles. Multicultural education is also concerned with the contribution of students towards effective social action.(banks ,2013)

2.Multicultural Based E-Learning

Al-Huwail, Al-Sharhan and Al-Hunaiyyan (2007) argued that, instructors often hear about the positive effects of e-learning systems that is being used somewhere and wonder if it would be useful in their own setting. When such a transfer of electronic learning occurs across different countries and cultures, there is a problem of portability. Further supporting evidence was presented by Gujar and Sonone (2004). They mentioned in their study that the adoption of educational and training to multicultural settings requires a new paradigm that includes an understanding of the deeper psychology of culture and the unique differences culture brings to a global workplace.

Culture is complex concept, and learning culture are affected by many variables. Moreover, the traditional, teacher-centered and content-driven learning culture does not necessarily produce the kind of learning that is need in the 21st century knowledge society. Striving for learning design that accommodates and accords with different students existing learning culture is therefore not only almost impossible, but also dangerous (Antal & Friedman 2004). Instead, it is crucial to develop learning designs that allow for dialogue, reflection and collaboration and thus creates a solid starting-point for the group to collaboratively create a multicultural, 21st century learning culture.

When moving out of the center of the learning process, the role of teacher changes and becomes more of a designer /script writer who delivers the pedagogical architecture for the " learning play" before it starts, and then acts as a participant, learning and facilitating in the network through the various movement among meta communicative levels in the networked dialogue(Sorensen,2007) Authentic learning programs must be implemented by using the very working method the students are expected to learn. In so doing in the cases described here, students gained significant and meaningful experience from the power of cooperation and feedback in learning.

To support authentic learning, interaction and collaboration within individual, pair and team is important. Furthermore, effective access to experts is also necessary. Also learning community is a significant thing.

Authentic e-learning seems to provide a useful framework for this type of a learning design. The elements of authentic e- learning all promote the type of activity that can lead to an increased cultural understanding and collaboration.

Design guidelines provide a framework for educators to create such authentic e-learning environment, specifically:

• Provide authentic contexts that reflect the way the knowledge will be used in real life.

• Provide authentic tasks and activities.

• Provide access to expert performances and the modeling of processes.

• Provide multiple roles and perspectives.

• Support collaborative construction of knowledge.

• Promote reflection to enable abstractions to be formed.

• Promote articulation to enable tacit knowl- edge to be made explicit.

• Provide coaching and scaffolding by the teacher at critical times.

• Provide for authentic assessment of learning within the tasks (Herrington et.al., 2010 p. 18).

Kerres and De Witt (2003) indicated that digital media will not substitute traditional approaches to learning and teaching as advocated by some e-learning enthusiasts a few years ago. Digital media do not question the existence of teachers or educational institutions as such and they will coexist with traditional approaches of teaching and training. In many cases, computer-based or internet-based trainings are accompanied by face-to-face (FTF) meetings to ensure the quality of learning and to reduce dropouts. The now widely adopted term 'blended learning' refers to all combinations of FTF learning with technology-based learning: traditional education can be enriched with the use of technology and learning with technology can profit from FTF meetings. The term blended learning, however, is still quite vague and does not provide a conceptual framework.

### 3. Learning Management System (LMS)

While the goal of a CMS is to store and distribute content, the goal of a learning management system (LMS) is to "simplify the administration of learning/training programs within an organization" (eLearning post, 2001, para. 3). LMS allow a learner to launch eLearning. LMS help to manage the interactions between the learner and the eLearning and other related resources. LMS help learners plan and monitor their progress in their learning journey. LMS are "software that automate[s] the administration of training events" (Hall, 2002b, p. 249). The automation of administrative functions via LMS can lead to considerable time, personnel, and resource savings. An LMS has significant administrative functions, which help an organization to "target, deliver, track, analyze, and report on….. Learning" (eLearning post, para. 3). These robust administrative functions enable organizations to track completion of mandated training (e.g. safety, hazardous materials), currency of professional certifications (e.g., continuing education units for medical and education professionals), and mandatory human resource related programs (e.g. sexual harassment, diversity) (Hall, 2002a, p. 5). LMS integrate tools to manage the tracking of learners and the content along with appropriate work flow processes. This combination of tools and processes allows an LMS to support the delivery and management of learning and tracking the results. As Robbins (2002) explains, learning management systems "enable companies to plan and track the learning needs and accomplishments of employees, customers, and partners" (p. 1).

Every LMS should have the ability to display a catalogue, register learners, track learner progress, and provide reports. LMS must "be capable of handling various delivery modes–online, instructor-led, self-paced, collaborative, facilitated, no facilitated, and the like"(Singh, 2001, p. 3).

Rosenberg (2001) emphasized that "A learning management system uses Internet technologies to manage the interaction between users and learning resources" (p. 161), which implies that these technologies are applied whether the LMS is operated internally or externally to a corporation. The goal of LMS is to manage processes related to delivery and administration of training and education. LMS are structured around the course rather than course content. Collaborative tools within LMS include capabilities that allow learners to work simultaneously with other learners using an internet/ intranet/extranet technology. *(Sonja Irlbeck, 2007)*

D. Methodology

Statistical Society Used in this study were Iranian and Iraqi students in MSC and PhD degree at Razi University of Iran .The study used latest version of open source software called Claroline. The experiment was done for a spring semester which was approximately 15 weeks in 2016.

As regards the research had an enrolment of 33 students that used Caroline, a total of 33 questionnaires were issued to the students for all question in 3 section for to answer the main question of the research. The justification for using this approach was that the population was small hence there was no need for sampling. The researcher issued out the questionnaires to participants during the last lecture of the semester. Then the students were given 60 minutes which was enough to complete the questionnaire. The approach was most effective because it yielded a high responds rate when compared to the approach where in the questionnaire are issued and students take them home and complete them before returning them at some future date. However, the draw back with this approach is that, although the responds rate is always high it is never 100 per cent because some questionnaires will not returned and some would be returned but will be spoiled. Some forms of spoil are, some questions may be not completed and also that some may have more than one answer in cases where they are required to tick just one box.

On the whole, the data were collected from answers of questions in all the questionnaires and were analyzed by SPSS with ANOVE and T-test. The data are normally distributed and the cronbakh test was used for real ability of this work. Descriptive statistics were calculated and were performed cross tabulations.

E. Analyze and Decision

The Iranian and the Iraqi students' satisfaction using the data contained in the 33 questionnaires completed by students and through trial software Mean difference using SPSS, the following results were obtained.

1. **Reliability**

Cronbach Alpha method to determine the reliability of the questionnaire for blended e–learning multicultural (Caroline) is followed by:

Table1.Cronbach Alpha method

**Case Processing Summary**

|  |  | N | % |
|---|---|---|---|
| Cases | Valid | 33 | 100.0 |
|  | Excluded[a] | 0 | .0 |
|  | Total | 33 | 100.0 |

a. Listwise deletion based on all variables in the procedure.

Table 1, indicates 33 participants in the test that were chosen as statistical population, so sample size equals to statistical population. The reliability of the questionnaire was done by Cronbach test. As shown in Table 2, Cronbach's alpha method is acceptable by 95% significance level.

Table2. Reliability Statistics

**Reliability Statistics**

| Cronbach's Alpha | Cronbach's Alpha Based on Standardized Items | N of Items |
|---|---|---|
| .000 | .000 | 10 |

2. **Satisfaction**

The students' satisfaction from multicultural blended e-learning (Caroline), was done by One-Sample T-Test And was compared to the obtained result average of the second part questions. The results were given in table3:

### Table.3. student's satisfaction from Claroline

**One-Sample Test**

| | Test Value = 0 | | | | | |
|---|---|---|---|---|---|---|
| | | | | | 95% Confidence Interval of the Difference | |
| | T | Df | Sig. (2-tailed) | Mean Difference | Lower | Upper |
| Question | 30.000 | 32 | .000 | 2.0000 | 2.000 | 2.000 |

The answer of students to the satisfaction questions of the questionnaires was in 95% of the significance level with possibility >5%. It was known that the satisfaction possibility (percent) of students from Caroline is 95%

### Table.4 satisfaction level from Claroline between the Iraqi and Iranian students

**Report**

| Nationality | | Q1 | Q2 | Q3 | Q4 | Q5 | Q6 | Q7 | Q8 | Q9 | Q10 |
|---|---|---|---|---|---|---|---|---|---|---|---|
| Iran | Mean | 2.00 | 4.00 | 2.00 | 1.00 | 2.00 | 2.00 | 2.00 | 2.00 | 2.00 | 2.00 |
| | Std. Deviation | .000 | .000 | 1.000 | .000 | .000 | 1.094 | .000 | 1.000 | .000 | 1.000 |
| | Std. Error of Mean | .091 | .062 | .000 | .000 | .000 | .000 | .000 | .000 | .000 | .000 |
| Iraq | Mean | 1.00 | 4.00 | 2.00 | 1.00 | 1.00 | 1.00 | 2.00 | 1.00 | 1.00 | 2.00 |
| | Std. Deviation | .000 | 1.000 | 1.000 | .000 | .000 | .000 | .000 | .000 | .000 | 1.064 |
| | Std. Error of Mean | .000 | .000 | .000 | .000 | .000 | .000 | .000 | .000 | .000 | .000 |
| Total | Mean | 1.00 | 4.00 | 2.00 | 1.00 | 1.00 | 2.09 | 2.00 | 1.00 | 2.09 | 2.00 |
| | Std. Deviation | .000 | .000 | 1.000 | .000 | .000 | .000 | .000 | 1.000 | .000 | 1.000 |
| | Std. Error of Mean | .088 | .000 | .000 | .087 | .000 | .000 | .000 | .000 | .000 | .000 |

The table 4 indicates that the satisfaction level from Claroline among the Iraqi students is more than the Iranian students. This issue is presented with significance level of each questions compared to the nationality in Std. Deviation row. By considering of significance level in table 4 .It was shown that the number of 0.0000 in Iraqi student's column is more than the other, so it was found that the significance level of the Iraqi students from Caroline is more than the Iranian students.

3. **Optimal Mix of Tools Data**

This section presents data on the mix of tools and proportions that make up blended multicultural-based e-learning. The data were collected using a questionnaire from the Iranian and the Iraqi students in the Razi University.

When it comes to the percentage that each multicultural element must constitute, the information is shown in chart 4. From that figure, it is evident that still pictures 19%, audio files 23%, video files 31% and text files 27%..

The mix tool presents of Claroline that consist of voice, Audio, Text, Still picture is illustrated in table 6.

**Table 5 mix tool presents of Caroline that consist of voice, Audio, Text, Still picture**

**One-Sample Statistics**

|  | N | Mean | Std. Deviation | Std. Error Mean |
|---|---|---|---|---|
| Video | 33 | 62.00 | 20.000 | 3.000 |
| Audio | 33 | 46.00 | 16.000 | 2.000 |
| Text | 33 | 53.00 | 19.000 | 3.000 |
| Still Pic | 33 | 37.00 | 14.094 | 2.000 |

According to the above table, 33 individuals were participated in this test (N=33). The average of mean error (Std. Error Mean) has been determined for each case of the test.

**Table 6. One sample Test and Mix tools**

**One-Sample Test**

| | Test Value = 33 | | | | | |
|---|---|---|---|---|---|---|
| | | | | Mean | 95% Confidence Interval of the Difference | |
| | T | Df | Sig. (2-tailed) | Difference | Lower | Upper |
| Video | 8.000 | 32 | .000 | 29.000 | 22.00 | 36.00 |
| Audio | 4.000 | 32 | .000 | 13.000 | 7.00 | 19.00 |
| Text | 6.000 | 32 | .000 | 20.000 | 13.00 | 27.00 |
| Still Pic | 1.000 | 32 | .071 | 4.000 | .00 | 9.00 |

In One-Sample Test Table 6 shows the high significant level for performance of the list items based on Two-sided test. It was known that 95% of the student participating in these questionnaire have mentioned to mix tools. This issue is due to the 95% significance level in Mix tools: Video, Audio, and Text.

*Figure 1:* The percentage of mix tools for blended multicultural eLearning (Caroline)

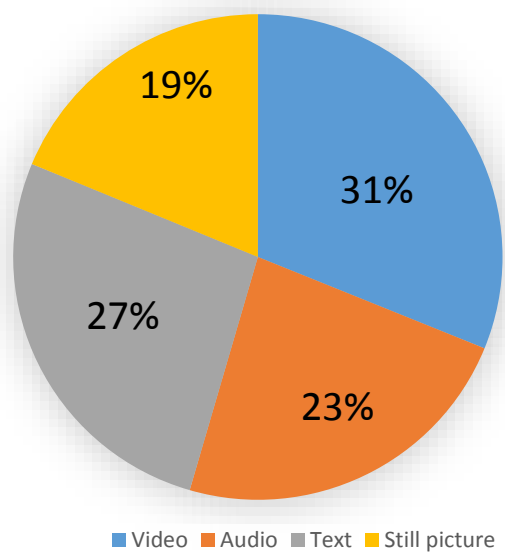

**Figure 1:Framework for Blended Multicultural-Based E-Learning**

The framework for blended e-learning presented in Figure 1 serves as a guide to plan, develop, deliver, manage, and evaluate blended learning programs. The framework for blended multicultural e-learning applications includes guidelines for selecting elements of a blended multicultural e-learning arrangement and the sequential ordering of these elements and their weight.

The major components that make up blended e-learning framework were identified as learners, instructors, smart classrooms, content and communication. From the questionnaires, issues that affect blended e-learning were identified. The issues were, interface design, institutional, resource support, technical, pedagogical, management, cultural, evaluation and ethical. The suitable frame for blended multimedia based on e-Learning is presented in Figure1.

In table 7 the most important parameters influenced in Caroline have been compared, The cases were analyzed by considering the most important parameters influence in the blended e learning that were answered by students.

Table 7. The most important parameters influenced in Caroline

**One-Sample Statistics**

|  | N | Mean | Std. Deviation | Std. Error Mean |
|---|---|---|---|---|
| Interface design | 33 | 7.00 | 1.000 | .000 |
| Evaluation | 33 | 7.00 | 1.000 | .000 |
| Resource support | 33 | 7.06 | 1.000 | .000 |
| Ethics | 33 | 7.00 | 1.000 | .000 |
| Technological | 33 | 7.00 | 1.000 | .000 |
| Pedagogical | 33 | 7.03 | 2.008 | .000 |
| Institutional | 33 | 7.06 | 1.000 | .000 |
| Culture | 33 | 7.00 | 1.000 | .000 |
| Management | 33 | 7.00 | 1.000 | .000 |

According to the above table, 33 individuals were participated in this test (N=33). The average of mean error (Std. Error Mean) has been determined for each case of the test.

Table 8. The parameter percentages influenced in Caroline

**One-Sample Test**

| | Test Value = 9 | | | | | |
|---|---|---|---|---|---|---|
| | | | | Mean | 95% Confidence Interval of the Difference | |
| | T | df | Sig. (2-tailed) | Difference | Lower | Upper |

| Interface design | -4.000 | 32 | .000 | -1.000 | -1.00 | .00 |
| Evaluation | -4.000 | 32 | .000 | -1.000 | -2.06 | .00 |
| Resource support | -7.000 | 32 | .000 | -1.000 | -2.00 | -1.00 |
| Ethics | -5.000 | 32 | .000 | -1.000 | -2.00 | .00 |
| Technological | -8.000 | 32 | .000 | -1.000 | -2.00 | -1.00 |
| Pedagogical | -5.000 | 32 | .000 | -1.000 | -2.00 | -1.00 |
| Institutional | -6.000 | 32 | .000 | -1.000 | -2.00 | -1.00 |
| Culture | -4.000 | 32 | .000 | -1.000 | -1.00 | .00 |
| Management | -5.000 | 32 | .000 | -1.000 | -1.00 | .00 |

One-Sample T Test, of 9 parameters of the questioner are the same and the Significance level of them is about 95%. It means that 95% of students participating in the comprehension test have the greatest impact on 9 parameters.

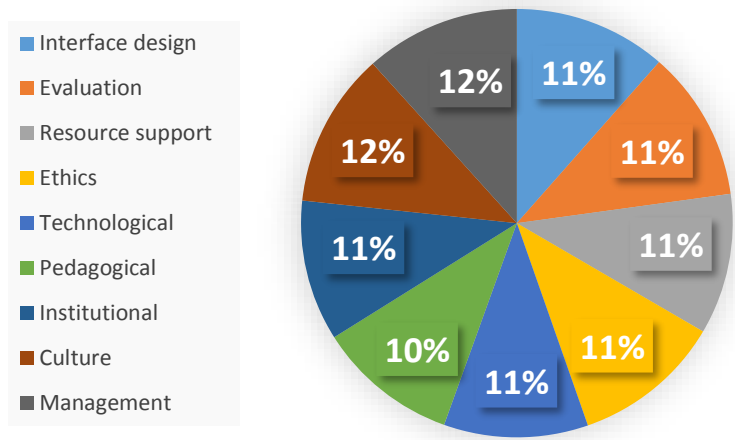

Figure 2. And the parameter percentages influenced in Caroline

Table 9 shows the results of analysis of the blended eLearning multicultural framework parameters compared base on the importance that was determined by students.

Table 9. The blended eLearning framework parameters

**One-Sample Statistics**

|  | N | Mean | Std. Deviation | Std. Error Mean |
|---|---|---|---|---|
| Communication | 33 | 7.00 | 1.000 | .000 |
| Smart class | 33 | 6.00 | 1.000 | .000 |

| | | | | | |
|---|---|---|---|---|---|
| Constructive Component | 33 | 6.00 | 1.000 | .000 | |
| Delivery Format | 33 | 6.00 | 1.000 | .000 | |
| Instructor | 33 | 7.00 | 1.000 | .000 | |
| Learner | 33 | 7.00 | 1.000 | .000 | |
| Content | 33 | 7.00 | 1.000 | .000 | |

According to the above table 9 individuals were participated in this test (N=33). The average of mean error (Std. Error Mean) has been determined for each case of the test.

In One-Sample Test 9 shows the high Significant level for the importance of e-learning framework parameters (Content, Learner, Delivery Format, and Communication) based on Two-sided test. It was known that 95% of the students participating in these questionnaire have mentioned the importance of e-learning in Content, Learner, Delivery Format, and Communication). This issue is due to the 95% significance level in these parameters.

Table 10. One-Sample Test for e-learning framework parameters

**One-Sample Test**

| | Test Value = 7 | | | | 95% Confidence Interval of the Difference | |
|---|---|---|---|---|---|---|
| | t | df | Sig. (2-tailed) | Mean Difference | Lower | Upper |
| Communication | .000 | 32 | .000 | .000 | .00 | .00 |
| Smart class | -1.000 | 32 | .081 | .000 | -1.00 | .08 |
| Constructive Component | -1.000 | 32 | .071 | .000 | -1.08 | .05 |
| Delivery Format | -1.000 | 32 | .000 | .000 | -1.09 | .00 |
| Instructor | .000 | 32 | 1.000 | .000 | .00 | .00 |
| Learner | .000 | 32 | .000 | .000 | .00 | .00 |
| Content | 1.000 | 32 | .000 | .000 | .00 | .00 |

Figure 3. The percent of the blended e-learning framework parameters

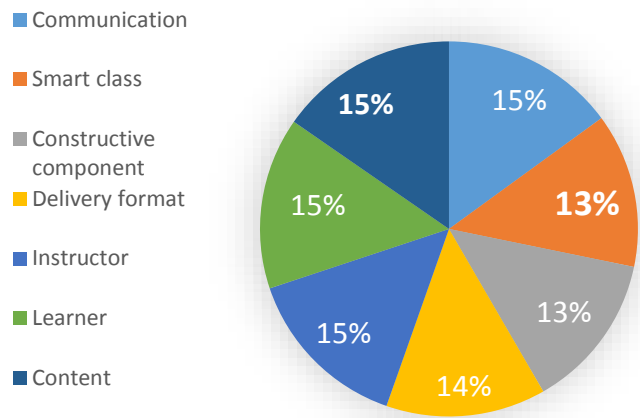

Figure 4 : Blended Multicultural-Based E-Learning Applications Framework

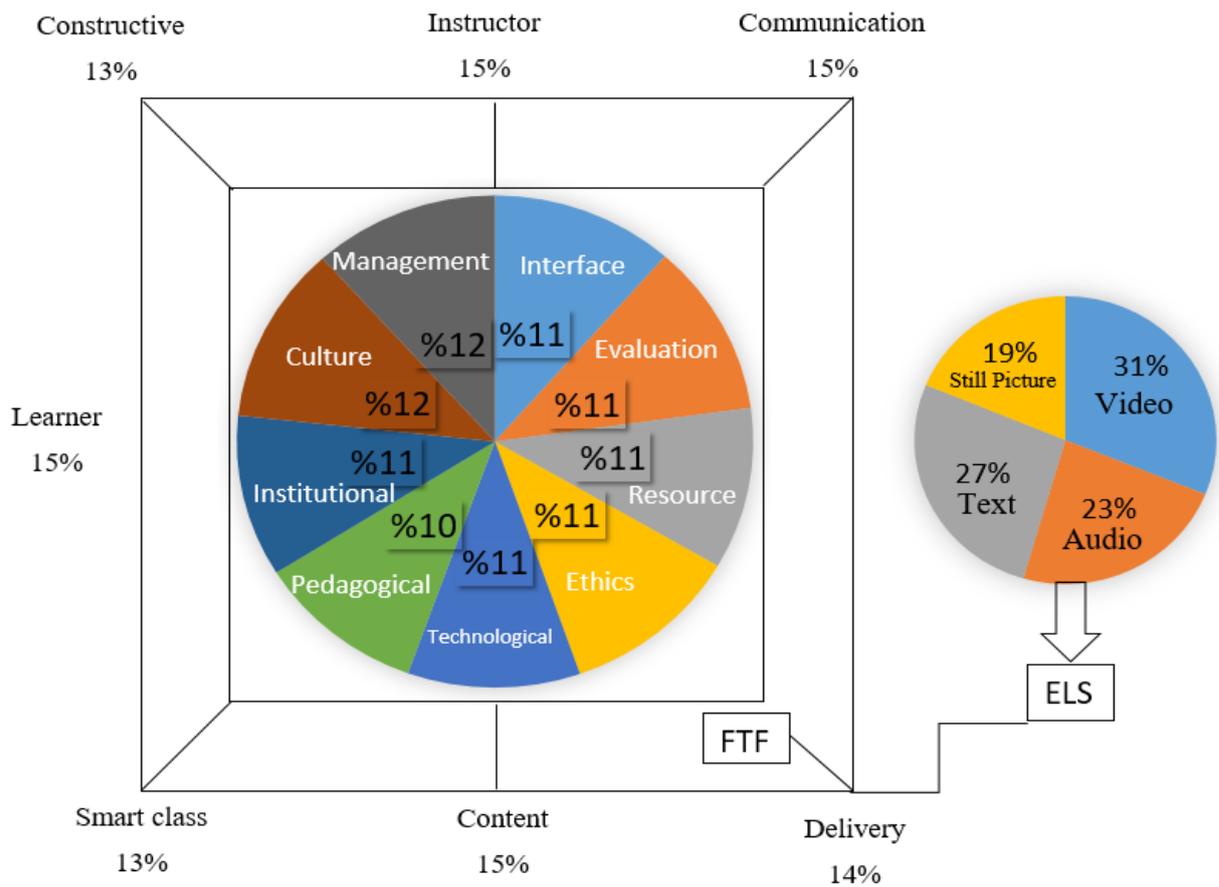

### F. Blended multicultural learning development steps

The steps which are involved in the development of blended multicultural e-learning are as follows:

1. Identifying all blended e-learning components and attaching a weight on each which shows the value of the component in the blended e-learning approach.
2. Naming all factors to be considered when establishing a blended e-learning approach. The issues are indicated in the framework in the figure 4.4. The weight of each factor in the framework is shown the framework. Here follows an explanation of the factors. The first factor is **institutional** in nature. Here the instructor must address issues such as organizational, administrative, academic affairs, student services, preparedness of the organization, availability of content and infrastructure. Also, the instructor must assess the feasibility of the institution to offer each learner the learning delivery mode independently as well as in a blended program. Lastly, the instructor must perform a needs analysis so as to understand all learners' needs. The second factor is **pedagogy.** Here the instructor must conduct content, learner, and learning objectives analyses. Also, the instructor must analyze the design and strategy aspect of e-learning. The third factor is **technology.** Here the instructor must create a learning environment and identify the tools required to deliver the learning program. Further, the instructor must establish the most suitable learning management system that manages multiple delivery types and the learning content management system that catalogs the actual content for learning program. Also, the instructor must acquire the server that supports the learning program and determine how the server will be accessed. Lastly, the instructor must address issues such as bandwidth and accessibility, security, other hardware, software, and infrastructure. The fourth factor is **interface design.** Here instructor must critically examine the user interface of each element in the blended learning program. Further, the instructor must analyze the usability of the user interface and ensure that the user interface supports all the elements of the blend. Lastly, the instructor must pay attention to issues like page and site design, content structure, navigation, graphics, and help. The fifth factor is **evaluation.** Here the instructor must evaluate the overall effectiveness of the learning program and must also evaluate the performance of each learner. The sixth factor is **Management.** Here the instructor must pays attention to blended learning administrative issues such as infrastructure and logistics to manage multiple delivery types. Also, the instructor must administer issues such as registration and notification, and scheduling of the different elements of the blend. The seventh factor is **resource support.** Here the instructor organize resources and make them available for learners. In addition, the instructor must make sure that they are always available either in person, via e-mail, or on a chat system. The eighth factor is **ethics.** Here the instructor must consider issues such as equal opportunity and nationality. The last factor is **culture.** Here the instructor must consider language, social, political, economic and religious issues.

3. Determining the best mix of to be used in presenting the materials content.
4. Explaining how the contents will be displayed. The instructor with cooperation of pedagogical expert determines how the contents will be displayed, the tools which will be used and the suitable content format, for example using PDF, Presentation, Text, HTTP, Wiki and XML.

5. Deciding the course activities: the lecturer with cooperation of pedagogical expert decides the suitable course activities, for example assignments, homework, chats, glossaries and quizzes.
6. Developing and designing. The lecturer develops the selected contents and activities and the multicultural technician designs the multicultural then put all in the target blended e-learning course.
7. Evaluating. The instructor insures that the blended e-learning course fits quality assurance procedures.
8. Updating. The lecturer with cooperation of the programmer makes the final modification upon the evaluation notes.
9. Allowing the blended e-learning course access.

Conclusions

This research focused on a suitable framework for blended multicultural e-learning applications. More specifically the research was so much about the impact that blended multicultural e-learning has on the level of satisfaction of learners needs. In addition the research also touched on the optimal mix of tools that go into blended e-learning. The paper began by presenting the problem statement, research objectives, hypothesis and other issues included in chapter one. Further, the research gave an outline of the literature on blended e-learning and in chapter three the basic concept and the research approach was presented. Lastly, chapter four presented research findings, analyzed and discussed them.

The major research conclusions were:

1. Blended multicultural e-learning brings maximum satisfaction to the learners' needs of students in developing country institutions such as the Razi University of Iran.

2. The optimal mix of blended learning is that still pictures must constitute 19% , audio files must constitute 23% , video files must constitute 31% and text files must constitute 27% .
3. The framework for blended multicultural e-learning is presented in chapter four.

Future research in this area is important to provide direction for the learning community to improve virtual learning experiences as well as for organizations to expand their reach and train global employees more efficiently. Finally, the researchers wish to emphasize the importance of this type of growth as, without it, we cannot hope to change the world through shared and consistent learning opportunities.

**Reference**

Al-Huwail N., Al-Sharhan S. and Al-Hunaiyyan A. 2007 Learning Design For A
   Successful Blended E-Learning Environment, Cultural Dimensions, Hawally,Kuwait.

Banks and Banks, eds. 2013. Multicultural Education, 'Multicultural Education: Characteristics and Goals', 'Culture, Teaching and Learning' (John Wiley & Sons)

Brusilovski, P. 2001 User modelling and user adaptation interaction, Adaptive
   hypermedia Vol. 11 (1/2), 87-110.

Chimombo J.P.G. 2005 Issues in basic education in developing countries: an exploration
   of policy options for improved delivery. CICE Hiroshima University, Journal of
   International Cooperation in Education, Vol. 8 (1), 129-152.

Gujar and Sonone 2004 E-Learning: The Modern Media, G. I. The changing face of
   design education. In World Conference on Educational Multimedia, Hypermedia
   and Telecommunications, volume 1, 2390–2396.

Heinze Aleksej 2008 Blended Learning: An Interpretive Action Research Study, Informatics Research Institute
   (IRIS), Salford Business School University of Salford, Salford, UK.

Itmazi Jamil A. and Tmeizeh Mahmoud J. 2008 Blended eLearning Approach for Traditional Palestinian Universities, IEEE Multidisciplinary Engineering Education Magazine, Vol. 3, No. 4, December 2008.

Kerres Michael and De Witt Claudia 2003 A Didactical Framework for the Design of Blended Learning Arrangements, Journal of Educational Media, Vol. 28, Nos. 2–3, October 2003, Carfax Publishing.

Kohn, T. *et al* 2008 Knowledge transfer with e-learning resources to developing countrie barriers and adaptive solutions. In: Breitner, M.H. et al. (eds.): E-Learning. Springer, Heidelberg.

Koper, E.J.R. 2003 Learning technologies: an integrated domain model. In W. Jochems, J. van Merrienboer

Rabahallah, K., & Ahmed-Ouamer, R. (2015, June). Creating e-learning web services towards reusability of functionalities in creating e-learning systems. In *Computer & Information Technology (GSCIT), 2015 Global Summit on* (pp. 1-6). IEEE.